\documentclass{sigchi-ext}
\usepackage[T1]{fontenc}
\usepackage{textcomp}
\usepackage[scaled=.92]{helvet} % for proper fonts
\usepackage{graphicx} % for EPS use the graphics package instead
\usepackage{balance}  % for useful for balancing the last columns
\usepackage{booktabs} % for pretty table rules
\usepackage{ccicons}  % for Creative Commons citation icons
\usepackage{ragged2e} % for tighter hyphenation
\usepackage{amsmath}
\usepackage{varwidth}
\usepackage{amsfonts}

 \usepackage{paralist}
\def\plaintitle{Safe Handover in Mixed-Initiative Control for Cyber-Physical Systems} 
\def\emptyauthor{}
\def\plainkeywords{criticality; mixed-initiative control; safety; cyber-physical system; car driving; multimodal interaction; human-computer interaction; natural language generation; description logics; automated planning; critical interaction; interdisciplinary}

\title{\plaintitle}

\numberofauthors{4}
\author{
   \alignauthor{
    {	
    }
    \textbf{Frederik Wiehr}\\
    \textbf{Anke Hirsch}\\
    \textbf{Florian Daiber}\\
    \textbf{Antonio Kr\"uger}\\ 
    \affaddr{German Research Center for Artificial Intelligence (DFKI)} \\
    \affaddr{Saarland Informatics Campus} \\
	}  \alignauthor{%
    \textbf{Alisa Kovtunova}\\
    \textbf{Stefan Borgwardt}\\
    \affaddr{Institute of Theoretical Computer Science, TU Dresden}\\
    \email{stefan.borgwardt@tu-dresden.de}
	} \vfil \alignauthor{%
    \textbf{Ernie Chang}\\
    \textbf{Vera Demberg}\\
    \affaddr{Saarland University}\\
    \affaddr{Saarland Informatics Campus} \\
    \email{vera@coli.uni-saarland.de} 
    } \alignauthor{%
    \textbf{Marcel Steinmetz}\\
     \textbf{J\"org Hoffmann}\\
    \affaddr{Saarland University}\\
    \affaddr{Saarland Informatics Campus} \\
    \email{hoffmann@cs.uni-saarland.de} 
    } 
 }

\definecolor{linkColor}{RGB}{6,125,233}
\hypersetup{%
  pdftitle={\plaintitle},
  pdfauthor={\emptyauthor},
  pdfkeywords={\plainkeywords},
  bookmarksnumbered,
  pdfstartview={FitH},
  colorlinks,
  citecolor=black,
  filecolor=black,
  linkcolor=black,
  urlcolor=linkColor,
  breaklinks=true,
}

\begin{document}

%% For the camera ready, use the commands provided by the ACM in the Permission Release Form.
\CopyrightYear{2020}
\setcopyright{rightsretained}
\conferenceinfo{CHI'20}{April  25--30, 2020, Honolulu, HI, USA}
\isbn{978-1-4503-6819-3/20/04}
\doi{https://doi.org/10.1145/3334480.XXXXXXX}
%% Then override the default copyright message with the \acmcopyright command.

\copyrightinfo{\acmcopyright}
\maketitle

% Uncomment to disable hyphenation (not recommended)
% https://twitter.com/anjirokhan/status/546046683331973120
\RaggedRight{} 

% Do not change the page size or page settings.
\begin{abstract}
For mixed-initiative control between cyber-physical systems (CPS) and its users, it is still an open question how machines can safely hand over control to humans. In this work, we propose a concept to provide technological support that uses formal methods from AI -- description logic (DL) and automated planning -- to predict more reliably when a hand-over is necessary, and to increase the advance notice for handovers by planning ahead of runtime. We combine this with methods from human-computer interaction (HCI) and natural language generation (NLG) to develop solutions for safe and smooth handovers and provide an example autonomous driving scenario. 
% TODO I would either use the term "highly automated driving scenario" or "autonomous driving scenario"!
A study design is proposed with the assessment of qualitative feedback, cognitive load and trust in automation.    

%The proposed concept has the goal to integrate HCI and formal methods, making human aspects of the human-machine system more accessible to formal analysis and thereby ensuring operational safety.<
  
\end{abstract}
%research questions the workshop addresses which we can stress on in the abstract
%How can non-expert users obtain an overall understanding of the reasoning of a system?
%Which styles of communication should be used to convey the state of a ubiquitous automated system?
%How to allow human interventions in complex automated procedures?
%To what extend should users be able to overrule the system behavior?
%How to design for an enjoyable interplay of non-expert users and automated systems?


% ACM Classfication

% TODO Why do we use the concept Haptic device?? I would rather suggest adding AI and NLG concepts!?!?



%\ccsdesc[300]{Human-centered computing~Haptic devices}

% Print the classficiation codes
%Please use the 2012 Classifiers and see this link to embed them in the text: \url{https://dl.acm.org/ccs/ccs_flat.cfm}

\section{Introduction}

Mixed-initiative control systems have shown that when decisions are made or suggested by automated systems it is essential that an explanation should be provided~\cite{DBLP:journals/aim/GoodmanF17}. %\textbf{[add reference here]}. 
An example of such a system is found in autonomous driving, where the cyber-physical system (CPS) derives responses from navigation functions based on human input. 
%Additionally, if the system needs to handover control to the human because it cannot handle a high-risk task autonomously, the user's attention needs to be re-directed to the driving task, and task-relevant information must be communicated. 
%It is thereby crucial for the machine to communicate the pertinent information in a concise, understandable and timely manner. 
%It will make the transition smoother, allowing for human corrective responses.

%side figure
\begin{marginfigure}
%\vspace{45pt}
  \begin{minipage}{\marginparwidth}
    \centering
    \includegraphics[width=1.0\marginparwidth]{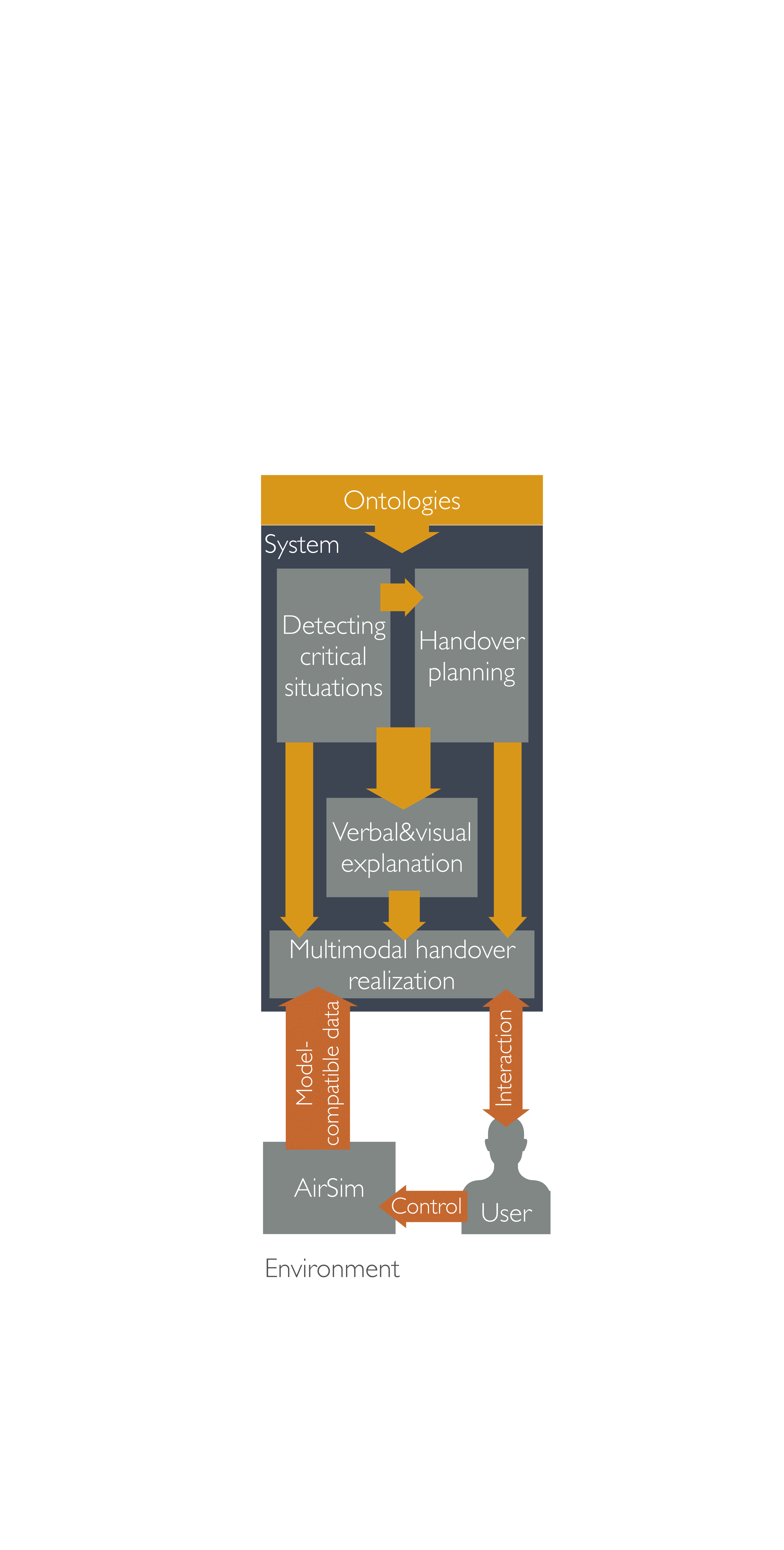}
    \caption{Proposed system architecture}~\label{fig:architecture}
  \end{minipage}
\end{marginfigure}

The purpose of this paper is to develop a framework to support a safe and stress-limiting handover of control to the driver, combining benefits derived from formal AI methods, human-computer interaction (HCI) and natural language generation (NLG).
The system must alert the user in the ``best possible way'' when a hand\-over is required. 
Practically, this poses the challenge that human factors may render the hand\-over process difficult to achieve~\cite{E2:endsley1996automation}, in part due to the lack of \emph{situation awareness} in the presence of a secondary task embedded in the experimental setups.
The driver, if distracted, must be able to promptly re-establish awareness of the situation, using sensory cues from the environment.  
The autonomous car can support this situation awareness by communicating its knowledge of the situation at the time of the transition. 
Moreover, the system may interact with the user, or takeover in the case of non-response.

Towards this end, we describe our implemented framework which utilizes the relevant environment-state information from the open-source simulator for the autonomous system AirSim \cite{shah2018airsim}, and propose a   
preliminary study design on the effect of varying the modality and timeliness of explanations in a simulated driving experience, with four handover situations.

\section{Related Work}
The related work in HCI can be separated into user modeling,
situation awareness/vigilance, and multimodal interaction. 

{\em User modeling} enables a system to maintain a conceptual understanding of the user (user
model \cite{E2:KW-89}, in which user differences need to be modeled
explicitly). 
%Examples of personalized applications range from museum guides \cite{E2:stock2007adaptive} to recommender systems \cite{E2:RRS-11}.
Tailoring hand-over requests by the system to individual users makes it possible to take into account the user's experience as well as individual differences relating to cognitive capacity.

\textit{Situation awareness} describes the human's awareness of the
environment, e.g.\ a critical situation for a task at hand  \cite{E2:endsley1996automation}. 
Such information depends strongly on the situation (e.g.~a pilot approaching an airport vs.\ a
driver navigating in dense traffic). 
Prior research utilizes operational quantitative human factors to assess the situation awareness of human operators \cite{E2:Adams1995}. 
%In particular, two approaches arise from different sensor placements:
%(1) body-worn sensors measure human activity
%directly \cite{E2:Lukowicz2004a}, while control
%elements such as the steering wheel of a car allow indirect assessment of human activities~\cite{E2:kass2007effects}.
%(2) Cognitive effects for assistive systems can be measured using self-assessments and questionnaires \cite{E2:kruger2004effects}, the Index of Cognitive Activity (ICA; \cite{E2:marshall2002index}), or
%employing the pupillometric measurement in dual task studies involving simultaneous language processing \cite{E2:autoui13,E2:demberg2016frequency}.
In highly automated systems, the risk of humans being
out-of-the-loop increases and thus a potential handover is more
difficult to achieve \cite{E2:endsley1996automation}. 
The degree of {\em vigilance} influences the ability of a human to attend to the
environment. 
Past research has developed vigilance measures based on
questionnaires, and sensors assessing heart rate,
eye movement and skin conductance \cite{E2:endsley1995measurement}.

{\em Multimodal interaction} has
the potential to increase the usability and thus the safety of
operation \cite{E2:cohen2004tangible}. 
It has been used for mobile applications and environments, including 
gesture and speech \cite{E2:wasinger2005integrating}, 
eye tracking and face detection \cite{E2:muller2009reflectivesigns}, and
tangible interaction \cite{E2:kalnikaite2011nudge} to adapt to the user's needs. 
Moreover, the styles of visualization can improve ``trust'' in autonomous
driving \cite{E2:pflegingetal17}, while identified feedback factors \cite{E2:koo2015did,E2:pflegingetal17} improve the understandability and trust of system decisions made in
autonomous driving.
%Our finding is that using a spatial
%representation of the environment to visualize the situation and the
%actions of the autonomous vehicle improved performance
%significantly. 

\section{Architecture}

\subsection{Planning}
% TODO citation missing!
The reasoning capabilities of modern AI
planners~\cite{ghallab2004planning,richter2010lama} provide the possibility to
foresee \emph{critical situations} the autonomous system's decisions may lead to
in the future, i.e.\ situations that cannot be reliably handled by the system on
its own.
Identifying such situations ahead of time is not only crucial for a successful
transfer of control to the user, but may actually allow the handover to be avoided
altogether in certain cases.

To identify and to anticipate critical situations, a planner builds upon an
abstract model consisting of two major components: 
(1) \emph{state features} that allow representation of an abstract view of the world for a specific point in time (e.g.\ current position and speed of the car), and the  (2) \emph{actions} the autonomous system can do at an abstract level (e.g.\  accelerating, changing lanes). 

In our architecture, AI planning is used for two purposes: (1) \emph{monitoring}, and (2)
\emph{replanning}. By default, in (1) the planner is only used to test the
autonomous system's decisions via simulations within the abstract world model;
and (2) uses planning to check the
existence of an alternative to the autonomous system's decisions in order to avoid other
critical situations.

\subsection{Description Logics}
Knowledge representation based on description logics (DLs) allows us to describe the complex environment in a so-called ontology, specifying constraints for the system states and %detecting inconsistencies, as well as 
for reasoning about the domain knowledge. Ontologies have, for example, been proposed for real-time patient monitoring~\cite{DBLP:journals/sensors/HristoskovaSZTT14,DBLP:conf/kes/RhayemMSG17}, 
%joerg: no need for this
%addressing security of the Internet of %Things~\cite{DBLP:journals/sensors/MozzaquatroAGMJ18}, 
detecting composite dance movements in annotated ballet videos~\cite{DBLP:conf/esws/RahebMRPI17}, and weather and turbine monitoring~\cite{DBLP:conf/aaai/BrandtKKRXZ17}.  

By design, description logics are close to human reasoning and can supply explanations for decisions made by a cyber-physical system based on an ontology. %Additionally, in order to model dynamic systems, description logics have extensions enabling temporal reasoning.
The DL component in our architecture provides input for the planning component.
% TODO "this route" has not been introduced before, or? 
By performing Boolean temporal query answering over a sequence of potential future states, the DL component assesses the criticality level of this route. Temporal queries describe different potentially dangerous situations on a road.
The level of criticality that is reported to the planning component depends on the number and severity of the situations that are detected. %When the value reaches a certain threshold, the planner either triggers replanning or initiates a handover.

\subsection{Natural Language Generation}

Description logic expressions serve as the semantic inputs to natural language generation (NLG) systems, which is analogous to the task of data-to-text generation.
Traditionally, this is dealt with using a pipeline consisting of \emph{content planning}, \emph{sentence planning} and \emph{linguistic realization}~\cite{reiter1997building}. 
While the format of the data varies from task to task, it typically involves the linearization step \cite{ws-2018-multilingual} where structured data are converted into sequences, before being processed by downstream systems for \emph{linguistic realization}.

% TODO "the handover message" ?
In the task of delivering handover message, NLG systems face a two-fold challenge: 
(1) there is insufficient annotated data for every situation,
(2) handover messages should not cognitively overload the user, but instead be at a situationally-appropriate level; see \cite{demberg2011linguistic,crocker2016information,hauser2019effects}. 
To this end, our proposed NLG system adopts (A) semi-supervised NLG techniques \cite{hong2019improving,chang2020unsupervised,shen2020neural,chang2020dart} that minimize the required data and 
(B) an automatic quality estimator \cite{de2018generating} that assesses the information density of the message. 

\subsection{Human-Computer Interaction}
For safe handovers in this context, the interaction side needs to be considered and adapted depending on the anticipated critical situation by the planner and DL component. Therefore, user modeling techniques can be applied to assess differences implied by the situation (such as cognitive load and vigilance: see below), as well as slowly changing individual user differences such as task familiarity (e.g.\ differentiating between novices and expert operators) \cite{E2:heckmann2003user}. 
To provide a basis for adaptations such as a multimodal hand-over realization, the HCI component of the system uses ontologies to provide access to and allow for interpretation of the user models~\cite{E2:heckmann2005gumo}. Although general user modeling is already established, the special safety-critical aspects of handover situations are still insufficiently addressed by prior research. Interface adaptations for cyber-physical systems have to be carefully integrated with the handover planning and DL components with a focus on safety critical situations. This includes developing concepts that respect cognitive implications of human operators.

\section{Conclusion \& Outlook}
\begin{marginfigure}
%\vspace{105pt}
  \begin{minipage}{\marginparwidth}
    \centering
    \includegraphics[width=1.0\marginparwidth]{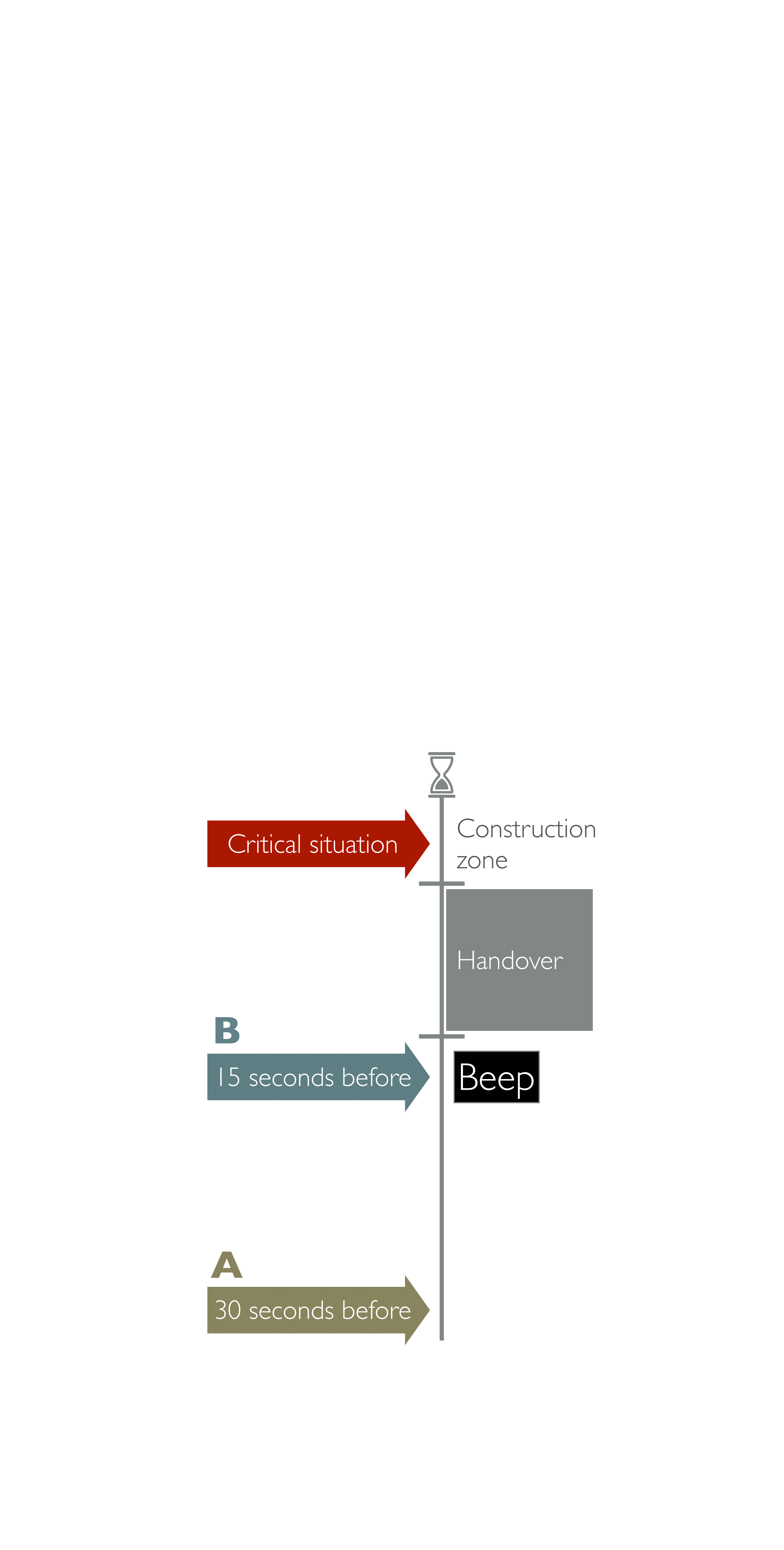}
    \caption{Handover Timing}~\label{fig:handover-timing}
  \end{minipage}
\end{marginfigure}

\marginpar{%
%\vspace{-45pt}
  %\fbox{%
    \begin{minipage}{\marginparwidth}
    \textbf{Acknowledgments} \\
    %\vspace{1pc}
    This work was partially supported by DFG grant 389792660 as part of TRR~248 (\url{https://perspicuous-computing.science}) and BMBF grant 01IW17004 (\url{http://tractat.dfki.de}).
    \end{minipage}
    %}
\label{sec:sidebar}
}

%With the aforementioned related work and approaches in mind, different experiments can be designed. 
% TODO inconsistent mix of present and future, i.e. sometimes presented as actual plan sometimes as should be done!
To gain insights into the use of a planner combined with description logics, we have chosen highly automated driving as a use case for handover in safety critical situations. Specifically, we plan to investigate handover scenarios where the driver has to take over control of the car in a driving simulator while performing a secondary task. The car or the systems can signal the handover in four different conditions. A goal here would be to test whether drivers would perceive additional verbal explanations as beneficial compared to a classical handover technique. The classic hand-over simply issues a notification at point B. Furthermore, the beep could be combined with either (a) a preceding request to take over (planner) or (b) a subsequent explanation about why the user had to take over (DL). The last condition could be a combination of planner and DL where the participants get an explanation and a request before the beep which signals them to take over. Planning enables the car to initiate the handover in time (at point A, see Figure~\ref{fig:handover-timing}) whereas the description logic alone can only generate the explanation in situ (at  B).
The criticality of the handover situation can be increased by approaching a construction zone in the driving simulator. Furthermore, the cause of an additional danger can be introduced by different driving scenarios of the cars in front of the participant. To avoid learning effects, four different driving scenarios should be combined with the four aforementioned conditions, like a unpredictable driver with odd steering behavior, a car that drives to the left lane, a very slow driver with sudden braking or a truck with an unsecured load.
Quantitative as well as qualitative data will be collected to conduct statistical analysis as well as participants' personal evaluation of each handover condition. The NASA-TLX \cite{hart1988development} and the Trust in Automation Questionnaire (TiA) \cite{korber2018theoretical} could assess cognitive load and trust in automated systems, respectively. Last, a semi-structured interview can be conducted to let the participants rank the different conditions and give reasons for the ranking.

A current design error in today's safety-critical systems is that these do not feature built-in concepts to pre-plan, or to recognize and explain problem causes to the user. Programs running in CPS participate in actions and decisions that affect humans, especially in highly automated vehicles, when a handover needs to be performed in critical situations. With the approach of combining formal methods with HCI, we believe that generating verbal and visual explanations in a timely manner using planning and description logics can ease the process for the user of regaining situational awareness and allow for a safe handover of control. We would like to further develop this idea during the workshop and discuss possible scenarios and experimental designs. 

\newpage

\section*{Acknowledgements}
This research was funded in part by the German Research Foundation (DFG) as part of SFB 248 ``Foundations of Perspicuous Software Systems''. We sincerely thank the anonymous reviewers for their insightful comments that helped us to improve this paper. 

%\section{Conclusion}
%Regaining situation awareness by a system request was assessed by measuring post-transition reaction time: more informative and advance notification should result in a faster reaction to a critical event. 

%\section{Acknowledgements}
%This work was supported by DFG grant 389792660 as part of TRR~248 (see \url{https://perspicuous-computing.science}).
%This work was partially supported by DFG grant 389792660 as part of TRR~248 (\url{https://perspicuous-computing.science}) and BMBF grant 01IW17004 (\url{http://tractat.dfki.de}).
\balance{} 

\bibliographystyle{SIGCHI-Reference-Format}
\bibliography{E2-situative-feedback-other,E2-situative-feedback-own,E2-situative-feedback,references}
\end{document}